# Topological Photonic Phase in Chiral Hyperbolic Metamaterials

Wenlong Gao[1,2*], Mark Lawrence[1*], Biao Yang[1], Fu Liu[3], Fengzhou Fang[2], Benjamin Béri[1], Jensen Li[1], Shuang Zhang[1†]

1. School of Physics & Astronomy, University of Birmingham, B15 2TT, UK
2. State Key Laboratory of Precision Measuring Technology and Instruments, Tianjin University, 300072, China
3. Department of Physics and Materials Science, City University of Hong Kong, Hong Kong, China

**Recently the possibility of achieving one-way backscatter immune transportation of light by mimicking the topological order present within certain solid state systems, such as topological insulators, has received much attention. Thus far however, demonstrations of non-trivial topology in photonics have relied on photonic crystals with precisely engineered lattice structures, periodic on the scale of the operational wavelength and composed of finely tuned, complex materials. Here we propose a novel effective medium approach towards achieving topologically protected photonic surface states robust against disorder on all length scales and for a wide range of material parameters. Remarkably, the non-trivial topology of our metamaterial design results from the Berry curvature arising from the transversality of electromagnetic waves in a homogeneous medium. Our investigation therefore acts to bridge the gap between the advancing field of topological band theory and classical optical phenomena such as the Spin Hall effect of light. The effective medium route to topological phases will pave the way for highly compact one-way transportation of electromagnetic waves in integrated photonic circuits.**

Despite the success of Landau's symmetry breaking theory in describing many different states of matter that exist in nature from melting points to superconductivity, recent developments in condensed matter physics have revealed a number of systems whose behaviour cannot be explained by symmetry breaking alone [1-9]. These are known as topologically ordered states as they remain invariant under smooth space-time transformations. Unlike normal media, the existence of boundary modes associated with topologically nontrivial systems depends entirely on the global properties of the bulk Hamiltonian regardless of the boundary formation. This is known as bulk-edge correspondence. As local perturbations to the edge cannot affect the properties of the bulk, these edge states are said to be topologically protected against disorder making them ideal candidates for overcoming dissipation and de-coherence in many quantum applications [1,2].

---

[*] These authors contributed equally to this work.
[†] s.zhang@bham.ac.uk

The Integer Quantum Hall Effect (IQHE) represents the most well understood form of topological order, both theoretically and experimentally [3]. By applying a strong magnetic field normal to a 2D electron gas, the bulk of the system becomes gapped due to the formation of topologically non-trivial Landau levels. One-way, scatter immune conduction channels are then guaranteed at the interface between this and a trivial insulating phase such as a vacuum, leading to a quantized Hall conductance. It has also been shown that topological protection can be achieved without breaking time-reversal symmetry in semiconductors with strong spin-orbit coupling – so called topological insulators [1,4,5]. In this case, the conduction channels appearing at the boundary are protected by time-reversal symmetry.

Topological features of gapless systems are also becoming subjects of intense study. The most dramatic prediction in this field is the appearance of surface Fermi arcs in so-called "topological semimetals". These are three-dimensional systems with multi-sheet Fermi surfaces, embracing momentum space analogues of magnetic monopoles. In this case bulk-boundary correspondence predicts surface states, which at the Fermi energy connect the bulk Fermi surface sheets in momentum space, thus forming unusual open momentum space curves ("arcs") instead of the familiar "Fermi circles" [6-8].

Recently, non-trivial topology in photonic systems has also been investigated fervently, driven by the similarity between non-interacting electron transport in crystals and photon transport in periodic dielectrics [10-16, 31]. A number of backscatter immune wave-guiding experiments have been proposed and demonstrated. The first of these involves the application of an external magnetic field to a 2D gyromagnetic photonic crystal, inducing chiral boundary modes at microwave frequencies as a result of time reversal symmetry breaking in analogy with the IQHE [10]. At higher frequencies where gyromagnetism is weak 2D lattices with bi-anisotropy, temporal modulation and ring resonators, in both strongly coupled periodic arrays and weakly coupled aperiodic arrays, have all been shown to exhibit topologically protected edge states at forbidden bulk band frequencies [12, 13, 31, 11]. By treating the distance $z$ propagated by a waveguide mode as a time-like coordinate, it has also been demonstrated that a honeycomb lattice of helical waveguides can be described as a Floquet topological insulator [14]. This system possesses robust one-way edge modes as long as $k_z$ is conserved by maintaining a fixed periodicity in the $z$ direction. Similarly, 2D planes in k-space have been shown to exist within 3D gyroidal photonic crystals which possess band gaps supporting topologically protected surface modes, analogous to Fermi arcs in Weyl semi-metals [15]. However, up until now investigations of bulk-edge correspondence, which is a key signature of topological phases, in photonic systems, have remained almost exclusively limited to lattice structures which depend on optical potentials varying spatially on the scale of the operational wavelength. A very recent theoretical work predicts that a magnetized zero-index-material can support scattering immune one way edge states [16], however, the generally very weak magneto-optic effect at optical frequencies limits the implementation of this scheme to very low frequencies.

In this paper we take a radically different approach towards topological photonics. Working in the long wavelength limit we show that topological behaviour can be observed in a time-reversal-invariant metamaterial, whose electromagnetic properties can simply be described by effective medium theory. By introducing chirality to a hyperbolic metamaterial we reveal that the polarisation degeneracy for waves propagating along the optical axis is lifted leaving behind topologically non-trivial equal frequency surfaces (EFS) separated in $k$-space, which is reminiscent of Fermi surfaces embracing momentum space monopoles. As expected, the interface between our metamaterial and a topologically trivial medium such as a vacuum or metal supports robust unidirectional surface states, serving as an optical analogue of Fermi arcs in topological semimetals. The topological protection of these surface states is highly robust, withstanding spatial deformations on all length scales as well as persisting for a wide range of electromagnetic parameters, including metamaterial structures with very low symmetry. Due to the simplicity and robustness of our design we believe this work to be of fundamental significance and may also be useful for the development of future photonic applications.

Over the past 10 years artificial periodic arrangements of subwavelength scattering elements known as metamaterials have gained a lot of attention from the scientific community. The reason for this interest is that metamaterials present a versatile platform for tailoring the material properties of a system allowing for the manipulation of electromagnetic waves in ways unseen in the natural world. Among the many unconventional electromagnetic properties available, it is now accepted that the strong inherent anisotropy associated with metamaterials can be very useful. Hyperbolic media represent one of the most extreme examples of anisotropy with permittivity and/or permeability tensors of mixed sign [17-20]. Due to the resulting mixture of metallic and dielectric behaviour, the propagation characteristics in hyperbolic systems can vary greatly depending on the direction of travel and polarisation of light. As the design of such material parameters is straightforward using effective medium theory [30], and these designs are relatively simple to fabricate, an array of exciting new phenomena have been demonstrated such as non-magnetic negative refraction [17,18] and enhanced Purcell effect [19]. In this paper we first consider a uniaxial hyperbolic metamaterial with permeability and permittivity,

$$\mu = \mu_r \mu_0, \quad \varepsilon = \varepsilon_0 \mathrm{diag}(\varepsilon_i, \varepsilon_i, \varepsilon_z)$$

where $\varepsilon_i > 0$ and $\varepsilon_z < 0$. Importantly, this system supports two propagating modes TE and TM, defined by $H_z = 0$ and $E_z = 0$ respectively. These modes are separated everywhere in $k$-space except for isolated degeneracies at $(k_x = k_y = 0)$, with a complementary existence in $k_z$ as shown in Fig. 1a. Based on previous investigations regarding degeneracies in electronic and photonic systems it is reasonable to ask whether or not this degeneracy can be lifted in such a way as to leave behind a topologically ordered set of EFS's. Below we show that this is indeed possible simply by adding chirality. Recent developments in metamaterials research have led to tremendous progress being made towards achieving extremely strong chirality at microwave, terahertz and even optical frequencies [21-27], paving the way for investigations of novel physics associated with strong optical activities. Here the incorporation of chirality within a hyperbolic medium causes the polarisation eigen-modes of the hyperbolic system to develop a sense of rotation with TE and TM like EFSs rotating in opposite directions, and being therefore topologically non-trivial with opposite topological invariants. This prediction is confirmed by the presence of protected unidirectional surface waves at the interface between an isotropic dielectric and the hyperbolic chiral medium proposed above.

In order to understand the topological behaviour of our metamaterial design we start by looking at the much simpler case of wave propagation in non-chiral isotropic media (Fig. 1b). It is well known that linearly polarised EM waves propagating in a twisted fibre will experience a rotation [28,29]. This rotation is caused by a difference in the Berry phase accumulated by circularly polarised modes with opposite helicities, originating from Maxwell's transversality conditions, $\boldsymbol{k}.\boldsymbol{D} = \boldsymbol{k}.\boldsymbol{B} = 0$, constraining the $\boldsymbol{D}$ and $\boldsymbol{B}$ fields to lie in the plane perpendicular to the wavevector. The existence of a relationship between the wavevector and the polarisation is analogous to spin orbit coupling experienced by electrons in certain semiconductors and is similarly responsible for the spin hall effect of light, otherwise known as the Imbert-Federov effect. Mathematically the Berry phase can be expressed as the integral of the Berry curvature, defined by $\boldsymbol{\Omega}_n = -i\boldsymbol{\nabla}_{\boldsymbol{k}} \times \langle U_n(\boldsymbol{k})|\boldsymbol{\nabla}_{\boldsymbol{k}}|U_n(\boldsymbol{k})\rangle$ where $U_n$ represents an Eigen-polarisation state, over the area $\Gamma_k$ bounded by a closed contour describing continuous changes in the propagation direction [9, 28],

$$\Phi_n = \iint_{\Gamma_k} \boldsymbol{\Omega}_n.\boldsymbol{ds}.$$

From this expression we can see that the phenomenon is purely geometrical, depending entirely on twists in the guiding medium which dictate the underlying path taken in $k$-space. As for Bloch bands or Fermi surfaces in a crystal, the integral of the Berry curvature over an entire EFS associated with a homogeneous photonic medium defines a topological invariant known as the Chern number [1, 7].

$$C = \frac{1}{2\pi}\iint_{\Pi_{surf}} \boldsymbol{\Omega}_n.\boldsymbol{ds}$$

For the case of scalar permittivity and permeability, $\boldsymbol{\Omega}_\sigma = \sigma\frac{\boldsymbol{k}}{k^3}$ (see supplementary material), where $\sigma = \pm 1$ corresponding to left and right handed circular polarisation [28].

Therefore, an isotropic photonic system must possess two $k$-spheres with $C = 2\sigma$. Despite the non-trivial topology of each circular polarised $k$-sphere, degeneracy causes the overall topological order to remain trivial. Nevertheless, the result $C = 2\sigma$ does lead to an exciting possibility. If these circularly polarised $k$-surfaces can be separated in such a way that maintains their non-trivial topology, the outcome should be a topologically ordered phase with protected surface states.

We now show how transforming an isotropic system to a hyperbolic chiral system results in topological order. The evolution of the EFSs is shown in Fig. 1b-f. We start by adding chirality $\gamma$ which is defined via the constitutive relations [23],

$$\boldsymbol{D} = \varepsilon\varepsilon_r\boldsymbol{E} - i\gamma\boldsymbol{H}/c$$

$$\boldsymbol{B} = \mu\mu_r\boldsymbol{H} + i\gamma\boldsymbol{E}/c$$

As circularly polarised light is always an eigen-mode of an isotropic medium, the addition of chirality changes the radius of the $k$-spheres (Fig. 1c), $k_\sigma = \omega(\sqrt{\mu_r\varepsilon_r} + \sigma\gamma)/c$, but cannot change their topology. Therefore, the Chern numbers associated with the outer and inner k-spheres are 2sgn($\gamma$) and -2sgn($\gamma$), respectively. By introducing anisotropy along the $z$ direction such that $\varepsilon_z > \varepsilon_i$, both k-spheres then become elliptically distorted, with the outer surface experiencing a larger deformation, i.e. the in-plane radius $k_r$ is seen to increase dramatically, as shown in Fig. 1d. After further increasing $\varepsilon_z$ towards infinity, $k_r$ of the outer k-surface becomes extremely large (Fig. 1e); While the Berry curvature distribution becomes complicated in such a highly anisotropic system, with increased concentration near $k_r = 0$ and decaying to zero for $k_z \to 0$, as long as the two EFSs do not intercept each other throughout this transformation, their topological identities, and therefore Chern numbers, must be preserved. Fig. 1f shows that finally pushing $\varepsilon_z$ through ∞ to negative values while keeping $\varepsilon_i$ fixed acts to transform the outer ($\sigma\gamma > 0$) $k$-surface from a closed surface to a slightly deformed two-sheeted hyperboloid, while the inner ($\sigma\gamma < 0$) $k$-surface once again remains closed. The Chern number of the central ($\sigma\gamma < 0$) k-surface is therefore unaltered with respect to the isotropic case, C=-2sgn($\gamma$), while the two hyperbolic sheets possess identical Chern numbers that sum to the opposite value, i.e., C=sgn($\gamma$) for each sheet. The chiral hyperbolic metamaterial thus displays three well separated and topologically non-trivial EFSs.

We now turn to studying how these topological features manifest themselves on the boundary of our metamaterial. In what follows we investigate systems with continuous translational invariance in the $z$ direction, thereby conserving $k_z$. The topologically non-trivial EFSs are expected to be joined by surface equi-frequency arcs, similar to Fermi arcs in topological semimetals [6, 8, 32-34]. To confirm this, here we solve for surface waves at the interface between our metamaterial design and a vacuum which involves searching for hybrid modes composed of both polarisation eigen-states that exist on each side of the boundary (see Supplementary Materials). In Fig.2a the results of this calculation reveal that between the EFSs any given surface can support just one propagating mode as expected. Importantly, the spatial separation of left and right moving surface waves at a certain $k_z$ prevents the occurrence of backscattering from any z-invariant disorder, as illustrated by Fig.2b.

This immunity to backscattering has also been confirmed using full wave simulations shown in Fig.3 in which a right moving surface wave propagates seamlessly around a large step defect. COMSOL is used to simulate the propagation of surface states at the interface between the topological metamaterial and vacuum. The simulation is performed in the x-y plane for different propagation constants $k_z$ in the gap between the bulk EFSs as indicated in Fig. 2a. As shown by Fig. 3a-c, the surface waves are not reflected or scattered by the presence of sharp corners. When the chirality of the metamaterial is flipped, the propagation of the surface wave is also switched to the opposite direction, as shown by Fig. 3d. These surface waves are therefore confirmed to be topologically protected.

While the system considered above is remarkably simple from a theoretical stand point, isotropic chirality to date presents some experimental challenges, especially at high frequencies. However, by extending the argument outlined in Fig.1 it is easy to show that a chirality tensor with just a single in-plane component, orthogonal to the optical axis of the permittivity, is required for topologically non-trivial EFSs to exist (Supplementary information). In addition, topological order persists for any hyperbolic permittivity tensor with only a single negative component, but this tensor does not have to be uniaxial (Supplementary information). The existence of this large topological parameter space has two consequences. Firstly, metamaterials with a lower degree of symmetry are much easier to design and fabricate making the experimental verification of our work not only possible but rather simple. To strengthen this claim an example of a realistic chirohyperbolic metamaterial structure for microwaves is outlined in the supplementary material. Secondly, the large range of material parameters that can be tolerated should lead to highly robust surface states that are immune to in-plane bulk material variations as well as surface defects.

To conclude, we have theoretically demonstrated that a topologically non-trivial phase can exist without a spatially dependent optical potential in homogeneous photonic systems with completely local material properties. Although ref [12] employs bianisotropic metamaterials to provide a spin orbit interaction within a 2d photonic crystal, where nonlocality and the band structure play important roles, the resultant topological phase relies crucially on the isolated four fold degeneracies that exist within permittivity and permeability matched photonic crystals with 3 fold rotational symmetry. In contrast, here we utilise an intrinsic form of spin orbit interaction originating from the transversality of Maxwell's equations, with our metamaterial parameters chosen simply to separate out particular polarisation states in $k$-space. Whilst achieving bulk propagating modes with non-trivial topology in photonic crystals requires complicated arrangements of complex materials, here the non-trivial topology of our metamaterial is controlled by only a few effective material parameters, namely the permittivity tensor and chirality. Importantly, hyperbolicity and chirality are responsible for protecting the edge states from backscattering and so time reversal symmetry does not need to be broken by using external magnetic fields, therefore, our design can be easily scaled to operate at any frequency. Since the building blocks of the topologically ordered metamaterial proposed here can be deep subwavelength, our work provides a platform for investigating highly confined topological surface states and manipulating surface waves with potentially subwavelength resolution. Our numerical investigation into a realistic metamaterial structure [supplementary material] clearly reveals topological behaviour in support of our initial effective medium approach; however, this is by no means the only or indeed optimal design for achieving topological order. In fact, the extremely large body of literature dealing with theoretical and experimental investigations of both hyperbolic metamaterials [17-20] and metamaterials with gigantic chirality [21-27], from microwave to optical frequencies, means that the experimental realization of metamaterial topological phases should be well within reach.

**Acknowledgment**: We thank Mike Gunn for stimulating discussions and feedback.

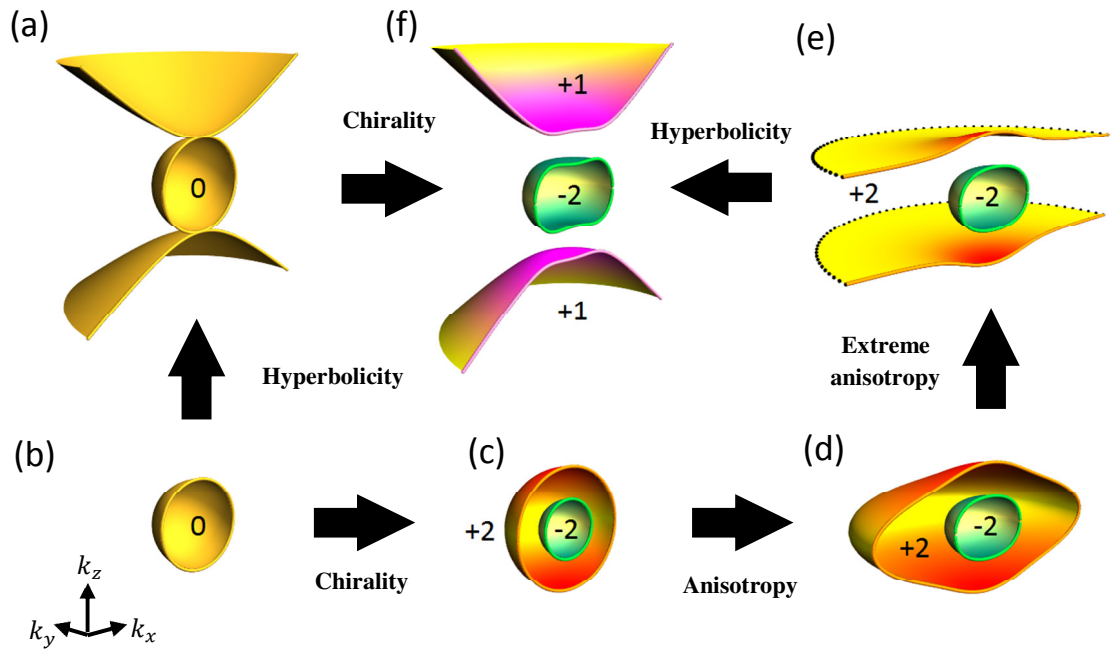

**Figure.1 Evolution of equi-frequency surfaces and their Chern numbers due to changing material parameters. a**, hyperbolic medium $\varepsilon_r$ = diag(4, 4, -3), $\gamma$ = 0. **b**, isotropic non-chiral medium $\varepsilon_r$ = 4, $\gamma$ = 0. **c**, isotropic chiral medium $\varepsilon_r$ = 4, $\gamma$ = 0.5. **d**, anisotropic chiral medium $\varepsilon_r$ = diag(4, 4, 30), $\gamma$ = 0.5. **e**, Highly anisotropic chiral medium $\varepsilon_r$ = diag(4, 4, ∞), $\gamma$ = 0.5, black dotted lines show the plot boundary of infinitely extended sheets. **f**, hyperbolic chiral medium $\varepsilon_r$ = diag(4, 4, -3), $\gamma$ = 0.5. $\mu_r$ is set to 0.5 for all the plots. The numeric label assigned to each surface represents its associated Chern number.

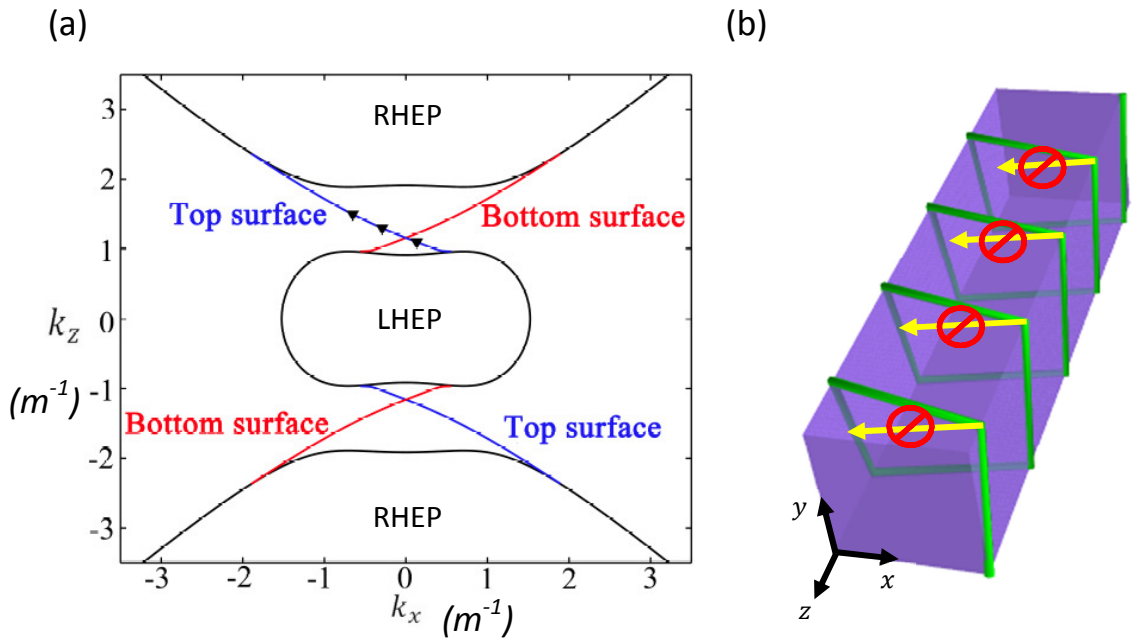

**Figure.2 Topologically protected surface states at the interface between a chiral hyperbolic metamaterial and a vacuum. a**, Volume (black) and Surface (red and blue) state dispersion for boundary in y-z plane with top (bottom) surface defined by metamaterial (vacuum) occupying the half space, $y < 0$. The parameters of the metamaterial are $\varepsilon_r$ = diag(4, 4, -3), $\gamma$ = 0.5. $\mu_r$ = 0.5; the black triangles represent the coordinates in k space corresponding to surface states in Fig. 3(a-c), LHEP and RHEP stand for Left and Right handed elliptical polarisation respectively. **b**, Chiral surface state propagating around a cylindrical metamaterial surrounded by air, despite the existence of sharp corners back-scattering is forbidden due to the absence of anticlockwise modes.

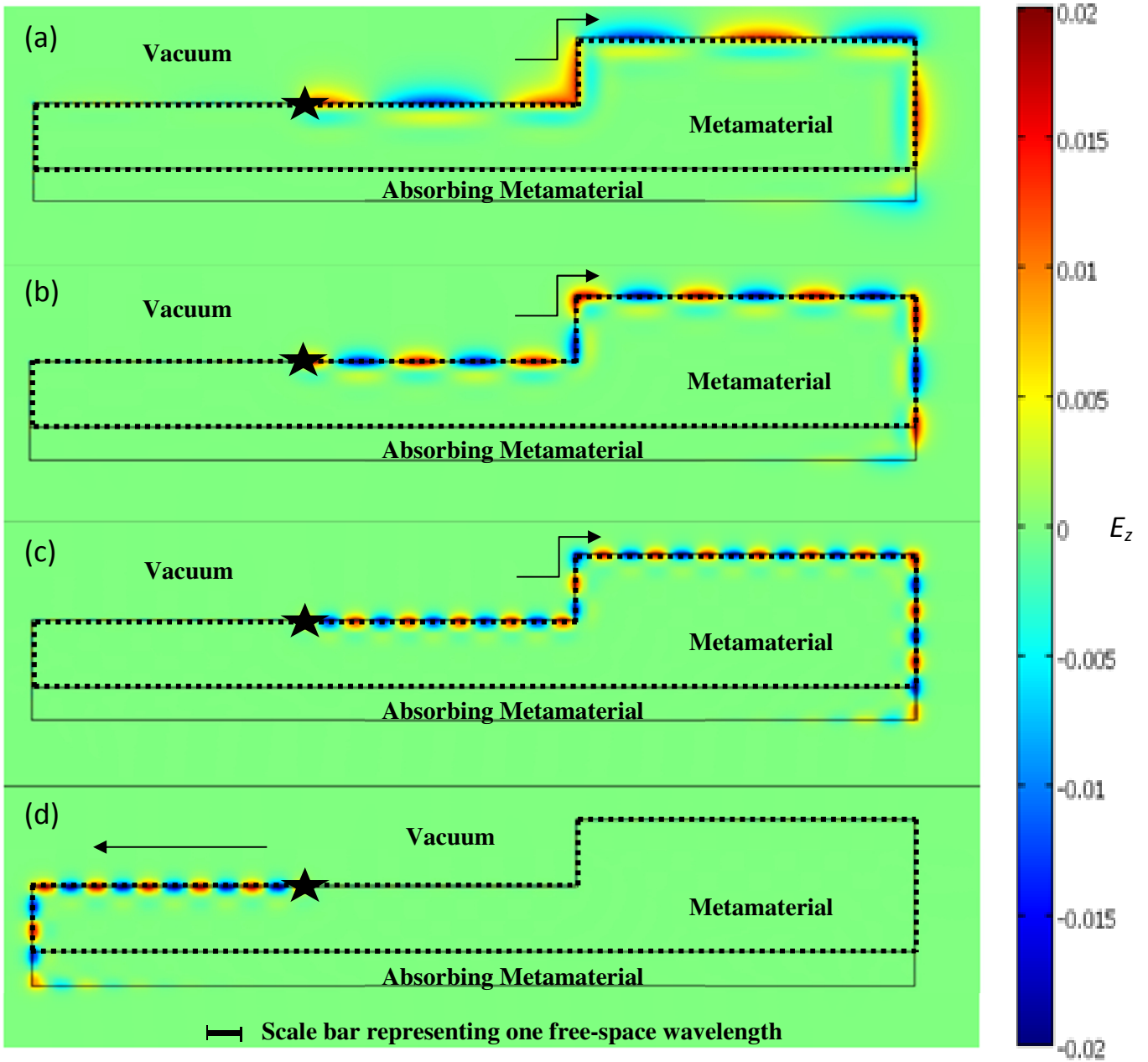

**Figure.3 Full wave simulations of topologically protected surface waves. a-c,** cross section view (x-y plane) of the field distribution at an interface between a hyperbolic chiral metamaterial and vacuum is shown with stars representing line sources with a z dependant phase gradient designed to excite electromagnetic waves with $k_z = 1.1k_0$ (**a**), $k_z = 1.3k_0$ (**b**) and $k_z = 1.5k_0$ (**c**), where $k_0$ is the free-space wavenumber. The metamaterial parameters are the same as in Fig. **2**. As $k_z$ is in the topological band gap a single surface wave can be seen propagating to the right moving smoothly through a step defect placed in its path due to the absence of left moving solutions. **d**, field distribution for a topological metamaterial with a negative chirality parameter $\gamma = -0.5$ at $k_z = 1.5k_0$. As the sign of the chirality parameter $\gamma$ is reversed the direction of propagation is switched to the left. In all simulations an absorbing layer with the same material properties as our topological metamaterial except for a large imaginary component added to the permittivity has been used to prevent the surface waves from interfering with themselves, allowing the unidirectionality of the boundary modes to be seen clearly.